\documentstyle[prc,aps,psfig]{revtex}
\begin{document}
\draft
\bibliographystyle{unsrt}
\title{$\rho$-mass Modification in $He^3$ - a Signal of Restoration of
Chiral Symmetry or Test for Nuclear Matter Models ?}
\author{Abhijit Bhattacharyya$^{1,a}$, Sanjay K. Ghosh $^{2,b}$
and Sibaji Raha$^{2,c}$}   
\address{1) Variable Energy Cyclotron Centre, 1/AF, Bidhannagar,
Calcutta 700 064, INDIA\\
2) Department of Physics, Bose Institute,
93/1 A.P.C.Road, Calcutta 700 009, INDIA}
\maketitle
\begin{abstract}
Two recent experiments have demonstrated that the effective $\rho$-mass in
nuclear medium, as extracted from the $^3He(\gamma, \pi^+ \pi^-)$ reaction, 
is substantially reduced. This has been advocated as an indication of 
partial restoration of chiral symmetry in nuclear matter. We show that 
even in the absence of chiral symmetry, effective mean field nuclear 
matter models can explain these findings quantitatively.
\end{abstract}

In-medium properties of hadrons is a field of high current
interest. In high energy heavy ion collisions, where
an environment of hot and dense hadronic matter is expected to be formed,
the modification of hadronic properties can indeed have important
physical consequences. The recent observation of enhanced dilepton
production in the low invariant mass domain in heavy ion 
collider experiments \cite{ceres,helios,na28} has triggered speculation
\cite{li} that the effective $\rho$-meson mass in the nuclear medium is 
decreased. Simultaneously, theoretical studies based on chiral 
perturbation theory ($\chi PT$) have led to the expectation that even at
finite densities (at or above nuclear density) there may be a partial
restoration of chiral symmetry, leading to the decrease of vector 
meson masses from their free values \cite{brown}. These questions 
assume great importance also in the context of quark-gluon plasma 
searches in heavy ion collisions, where the hadronic matter constitute
the background to the sought-for signals and thus must be controlled to
a high degree of accuracy before conclusive evidence for the quark-gluon
plasma can be extracted from the experimental data \cite{alam}.
\par
The evidence (or more appropriately, indication) of the $\rho$ mass 
renormalization, referred to above, is indirect. Very recently, however,
a direct measurement of the invariant $\rho$ mass in photoproduction of
$\rho^0$ on $He^3$ has been reported in the literature \cite{lolos,huber}.
The decrease in the $\rho$ mass found by these authors is quite substantial
($\delta m_{\rho_0} \sim 280 \pm 40 $ MeV), so much so that it has been 
argued \cite{lolos} that such large decrease cannot be explained by the 
mean field picture of nuclear matter \cite{h}. These authors suggest 
that this should be taken as a signature of (partial) restoration of chiral 
symmetry \cite{brown1} in ground state nuclei. In this work, we show that 
such a conclusion is premature since a proper inclusion of the relevant 
interactions in a mean field description has the effect of reducing the
$\rho$-mass to the desired level.
\par
The most popular mean field model for nuclear matter is the Walecka model 
\cite{d}, which was proposed first in 1974 and has been greatly modified 
over the years by a number of authors; for a recent comparative study among
the various versions, see \cite{e}. This serves as the prototype of most
effective mean field theories of nuclear matter and for the present purpose,
we concentrate our attention only to the Walecka model, the basic Lagrangian
for which is given by \cite{e1}:

\begin{eqnarray}
{\cal L} = {\bar \psi} (i \gamma_\mu \partial^\mu  - m_n) \psi 
- \left[g_\omega {\bar \psi} 
\gamma_\mu \psi \omega^\mu - {1 \over 4}G_{\mu 
\nu} G^{\mu \nu}
+ {1 \over 2} m_\omega^2 \omega_\mu \omega^\mu \right]  \nonumber\\
+ {1 \over 2}(\partial_\mu \sigma \partial^\mu \sigma - m_\sigma^2 
\sigma^2) +g_\sigma {\bar \psi} \sigma \psi 
- {1 \over 4} F_{\mu \nu} F^{\mu \nu} \nonumber\\
+ {1 \over 2} m_\rho^2
\rho^2 
+ g_\rho {\bar \psi} \gamma^\mu {\bf \tau.}\psi \rho_\mu 
+ f_\rho {\bar \psi} \sigma^{\mu \nu} {\bf \tau.} \psi {{\partial_\mu} 
\over {2m}} \rho_\nu
\label{eq:lag}
\end{eqnarray}
 
In the above equation $\psi$, $\sigma$, $\omega^\mu$ and $\rho^\mu$ 
are, respectively, the nucleon, the $\sigma$, the $\omega$ and the 
$\rho$ meson fields; $m_n, m_\sigma$, $m_\omega$ and $m_\rho$ are the 
corresponding masses; $g_\sigma$ and $g_\omega$ are the couplings of the 
nucleon to $\sigma$ and $\omega$ mesons, respectively; $g_\rho$ and 
$f_\rho$ are the vector and tensor couplings of rho meson;  
$G_{\mu \nu} = \partial_\mu \omega_\nu - \partial_
\nu \omega_\mu$ and $F_{\mu \nu} = \partial_\mu \rho_\nu - \partial_
\nu \rho_\mu + ig_\rho \left[\rho_\mu,\rho_\nu\right]$.

In general, for medium and heavy nuclei one uses the mean field
approximation (MFA) and then a set of
coupled differential equations is solved self-consistently to get the
field values as a function of $r$ \cite{h0}. But for a light nucleus like
${He}^3$, the MFA may not be reliable. Therefore, we have used a simple
approach, {\it a la} Saito {\it et al.} \cite{h}, to
calculate the effective $\rho$ mass in helium.

In this paper we have used a simple Gaussian form for the density
distribution of ${He}^3$, in which the width parameter $\beta_3$ is
fitted to reproduce the rms charge radius of ${He}^3$ {\it i.e.}
1.88 fm \cite{h}. The density profile is given in figure 1.
So once we know the density distribution of ${He}^3$, one can easily
calculate the effective $\rho$ mass, ${m_{\rho}}{^*}$, as a function of
radius, all the fields being known as a function of baryon density. 

The $\rho$ mass has been calculated from eq. (\ref{eq:lag}),
in the usual manner, at the one loop level. The expression for the
$\rho$-mass is given by
\begin{equation}
m_\rho^{*2} = m_\rho^2 + \Pi_{vac} + \Pi_{med}
\end{equation}
where
\begin{eqnarray}
\Pi_{med} = \sum_{B = n,p} {{8g_\rho^2} \over \pi^2} \int_0^{k_{FB}} 
{{p^2 dp} \over {E_p \left(m_\rho^{*2} - 4E_p^2\right)}} \nonumber\\ 
\times \left[{2 \over 3} \left(2p^2 + 3m_n^{*2}\right) 
+ m_\rho^{*2} \left\{2m_n^*\left({{c_\rho} \over {2m_n}}\right) 
\right. \right. \nonumber\\
\left. \left. - {2 \over 3} \left({{c_\rho} \over {2m_n}}\right)^2 \left(p^2 + 3m_n^{*2}\right)
\right\}\right]
\end{eqnarray}
\vskip 0.01in
\begin{eqnarray}
\Pi_{vac} = {{g_\rho^2} \over \pi^2} m_\rho^{*2} \left[I_1 + 
m_n^* \left({{c_\rho} \over {2m_n}}\right) I_2 \right. \nonumber\\
\left. + {1 \over 2} \left({{c_\rho} \over {2m_n}}\right)^2 \left(m_\rho^{*2} I_1 + 
m_n^{*2} I_2 \right) \right]
\end{eqnarray}
\vskip 0.01in
\begin{equation}
I_1 = \int_0^1 dx x(1-x) ln\left[{{m_n^{*2} - m_\rho^{*2} x(1-x)} \over 
{m_n^2 - m_\rho^2 x(1-x)}}\right] 
\end{equation}
\vskip 0.01in
\begin{equation}
I_2 = \int_0^1 ln\left[{{m_n^{*2} - m_\rho^{*2} x(1-x)} \over 
{m_n^2 - m_\rho^2 x(1-x)}}\right] 
\end{equation}
In the above set of equations $c_\rho \equiv f_\rho /g_\rho$. 
\par
There are two coupling constants involved in the above set of equations. 
One is the vector coupling of the $\rho$-meson $g_\rho$ and the other 
is the tensor coupling $f_\rho$ (or equivalently $c_\rho$). This is where 
the difference between our approach and the earlier works arises; previous 
authors \cite{h} neglected the tensor coupling of the $\rho$ to the 
nucleon. We have used three sets of coupling constants, shown in table 1. 
The density dependence of the $\rho$-meson mass, for these three sets 
of parameters, has been shown in figure 2.

\begin{table}
\begin{center}
\begin{tabular}{|lcc|}
\hline
                    &        &          \\
               model&$g_\rho$&$c_\rho$  \\
                    &        &          \\
\hline
                              &         &       \\
                Bonn Potential \cite{f}& $2.63$  &$6.1$ \\
                QCD Sum rule \cite{g} & $2.5 \pm 0.2$& $8.0 \pm 2.0$ \\
                Walecka Model & $8.912$  &$6.1$ \\
\hline
\end{tabular}
\end{center}
\caption{ Parametr values for different models.} 
\end{table}
In order to compare the results for the effective $\rho$ mass with the 
experimental values, we calculate the average mass of the $\rho$-meson in 
the $He^3$ nucleus. The average mass is defined as 
\begin{equation}
\langle m_\rho^* \rangle = {{\int d^3r m_\rho^* (r) \rho_B (r)} 
\over {\int d^3r \rho_B (r)}}
\end{equation}
In table 2, we show the average $\rho$-mass for the different sets
of parameters.

\begin{table}
\begin{center}
\begin{tabular}{|lc|}
\hline
                    &                  \\
               Model&Average Mass  \\
                    & $(MeV)$          \\
                    &           \\
\hline
                              &               \\
                Bonn Potential\cite{f}& $536$   \\
                QCD Sum rule\cite{g}  & $449 - 565$ \\
                Walecka Model & $304$   \\
\hline
\end{tabular}
\end{center}
\caption{ Average mass of $\rho$ meson for different models.} 
\end{table}

There have been two recent papers on the density variation of $\rho$-mass 
inside the $He^3$ nucleus. Both of them are the results from the $\rho^0$ 
photoproduction experiment of $He^3$. The first one is in the energy range 
$E_\gamma = 800 - 1120 MeV$ and the second for 
$E_\gamma = 380 - 700 MeV$. The first paper finds a drop in the 
$\rho$-mass of $160 \pm 35 MeV$, {\it i.e.} $m_\rho^*$ is in the 
range $575-645 MeV$. The other study finds an effective $\rho$-mass in 
the range $450-530 MeV$. 

In almost all the previous studies of the $\rho$-meson inside a light 
nucleus from the mean field approach, the tensor coupling of the
$\rho$-meson to the nucleon \cite{h0,h} was not
included, as already mentioned. As a result, the variation of $\rho$-mass
was rather soft in all the previous cases. Here, the
incorporation of the tensor coupling leads to a change in the $\rho$-meson
mass which is much larger and we get results which are very close to the
experimental findings. For example, the Bonn potential parameter set
\cite{f} yields $\langle m_\rho^* \rangle = 536 MeV$.  On the other
hand, for the QCD sum rule case \cite {g}, we get
$\langle m_\rho^* \rangle = 449 - 565 MeV$. For the Walecka model parameter
set, the value of $\langle m_\rho^* \rangle$ is somewhat lower.

On the basis of above observations we argue that the reduction of the
effective mass of $\rho$- meson in $He^{3}$ need not be an unambiguous
signal for the restoration of chiral symmetry, as suggested by the authors of
ref. \cite{huber}. In particular, even the mean field model of nuclear matter
is capable of accommodating such substantial changes in the effective
$\rho$ mass, if all the interactions are properly taken into account.

We would like to mention here that the present calculation
is really an estimate of the behaviour of $\rho$ meson mass inside a
nucleus. To get a quantitative estimate and compare with the experiments
mentioned above, one should do a full calculation of photoproduction
processes. Such a non-trivial calculation is presently under study.

The work of SKG has been supported in part by the Council of Scientific
and Industrial Research, Govt. of India.

\newpage
\begin{figure}[htb]
\psfig{file=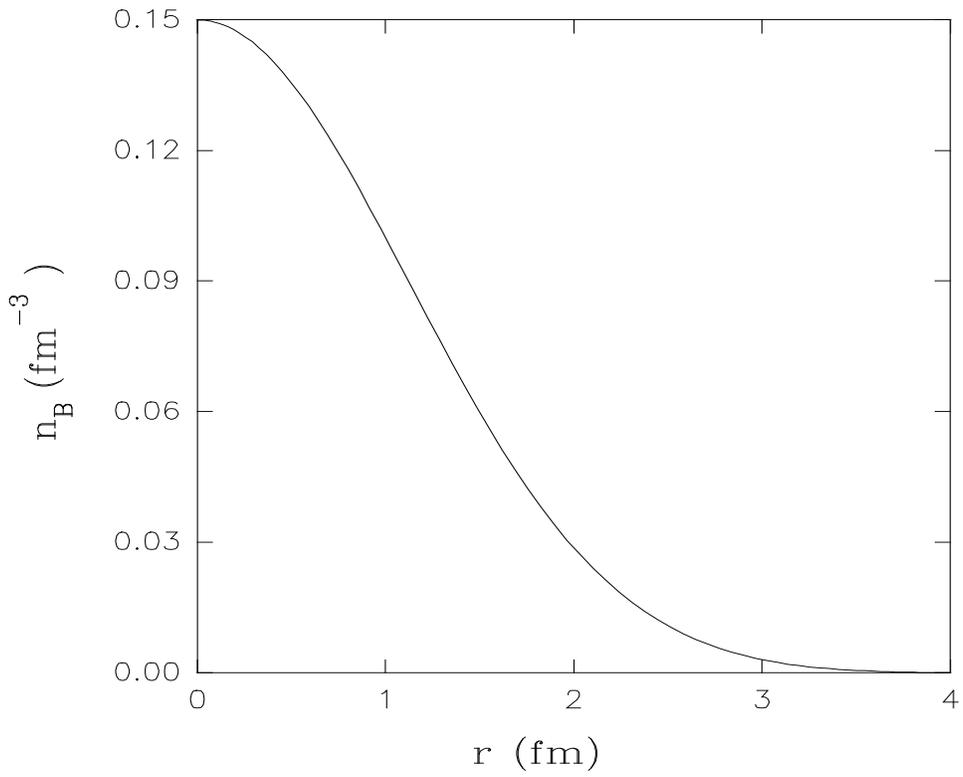,width=5in,height=4in}
\caption{Density profile of $He^3$}
\end{figure}
\begin{figure}[htb]
\psfig{file=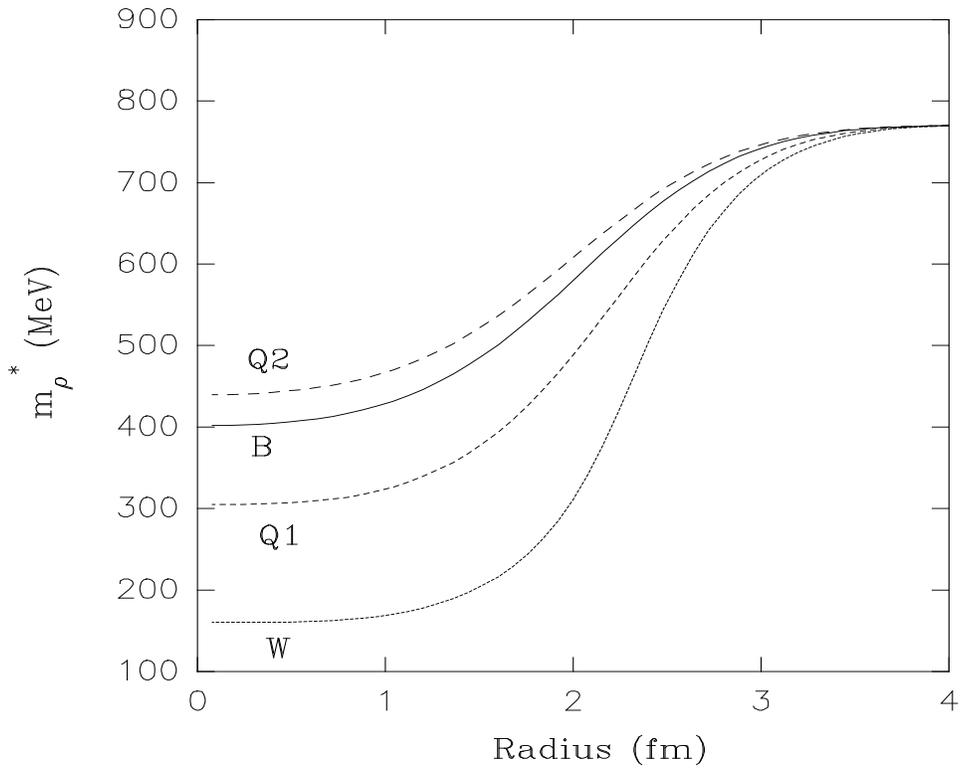,width=5in,height=4in}
\caption{Density dependence of rho mass, Q1 and Q2 are the upper and lower
mass limits for the QCD sum rule parameter set, B for Bonn potential
parameters and W is for Walecka model parameters.}
\end{figure}

\end{document}